\documentclass[preprint,10pt]{elsarticle}

\usepackage{graphicx}
\usepackage{amsmath, amssymb, amsthm,mathtools, stmaryrd, bm, gensymb}
\usepackage{url}
\usepackage{tikz, pgfplots}
\usepackage{siunitx}
\usepackage{algorithm, algpseudocode}
\usepackage{siunitx}
\usepackage{optidef}
\usepackage{siunitx}

\RequirePackage[T1]{fontenc}
\RequirePackage[utf8]{inputenc}

\usetikzlibrary{plotmarks,fit,shapes}

\newcommand{\esp}[1]{\ensuremath{\mathbb{E}\left[#1\right]}}

\newcommand{\B}[1]{\mathbf{#1}}

\newcommand{\R}[1]{\mathrm{#1}}

\newcommand{\bof}[1]{{\mbox{\boldmath$#1$}}}

\journal{Signal Processing}

\begin{document}

\begin{frontmatter}
\title{Channel Parameter Estimation for Millimeter-Wave Cellular Systems with Hybrid Beamforming}

\author[add1]{Fazal-E-Asim\corref{cor1}}
\ead{fazalasim@gtel.ufc.br}
\author[add3]{Felix Antreich}
\ead{antreich@ieee.org}
\author[add1]{Charles C. Cavalcante}
\ead{charles@gtel.ufc.br}
\author[add1]{Andr\'e L. F. de Almeida}
\ead{andre@gtel.ufc.br}
\author[add1,add2]{Josef A. Nossek}
\ead{josef.a.nossek@gtel.ufc.br}
\address[add1]{Wireless Telecommunications Research Group (GTEL), Universidade Federal do Cear\'a (UFC), Campus do Pici, s/n, Bloco 722, CP 6005, 60455-760, Fortaleza, Brazil}
\address[add2]{Department of Electrical and Computer Engineering, Technical University of Munich (TUM), Germany}
\address[add3]{Department of Telecommunications, Aeronautics Institute of Technology (ITA), Brazil.}

\cortext[cor1]{Corresponding author}
\tnotetext[t1]{The authors would like to thank Coordenac\~ao de Aperfeicoamento de Pessoal de Nivel Superior - Brasil (CAPES) - Finance Code 001, and CNPq (Procs. 309472/2017-2 and 306616/2016-5) for the partial financial support. }

\begin{abstract}
    To achieve high data rates defined in 5G, the use of millimeter-waves and massive-MIMO are indispensable. To benefit from these technologies, an accurate estimation of the channel parameters is crucial. We propose a novel two-stage algorithm for channel parameters estimation. In the first stage, coarse estimation is accomplished by applying parameter estimation via interpolation based on a DFT grid (PREIDG) with a fixed look-up table (LUT), while the second stage refines the estimates by means of the space-alternating generalized expectation maximization (SAGE) algorithm. The two-stage algorithm uses discrete Fourier transform beamforming vectors which are efficiently implemented by a Butler matrix in the analog domain.  We found that this methodology improves the estimates compared to the auxiliary beam pair (ABP) method. The two-stage algorithm shows efficient performance in the low signal to noise ratio regime for the channel parameters i.e. angles of departure, complex path gains and delays of the multipaths. Finally, we derived the Cram\'er-Rao lower bound (CRLB) to assess the performance of our two-stage estimation algorithm.
\end{abstract}


\end{frontmatter}

\section{Introduction}
Millimeter-Waves and massive multiple-input-multiple-output (MIMO) are the two essential candidates to achieve the high promising data rates for the fifth generation (5G) cellular systems \cite{Zpi_2011,Rappaport_2013}. On one hand, introducing high frequencies will facilitate the introduction of large antenna arrays at the base station (BS), which will end up with high beamforming gains but on the other hand, the large bandwidth in mmWaves with large antenna arrays impose the challenge of hardware implementation \cite{FRusek_2013}. Therefore, energy efficiency with addition to spectral efficiency becomes an important design goal.
\\
In order to fully exploit the benefit of using large antenna arrays both at the transmitter and receiver, full digital precoding is used, which means all signal processing is done in the baseband, introducing the cost of one radio frequency (RF)-chain per antenna, which turns out with high implementation complexity in addition to power consumption. Multiple solutions are proposed to cater for this challenge including less number of RF-chains termed as hybrid beamforming \cite{CLeeWChung_rf_2017,JZhuWXu_rf_2017,LYangYZeng_rf_2016,LLianALiu_rf_2017} and low resolution analog-to-digital converter (ADC)s \cite{AMezghaniJANossek_adc_2008,AMezghaniJANossek_adc_2007,JMoRWHeath_adc_2015}.
\\
Another practical example in mobile communication is hybrid beamforming designed to overcome the implementation and energy challenges in mmWave massive-MIMO precoding. In hybrid beamforming, a small number of RF-chains are introduced to excite a large number of antenna elements both at the transmitter and receiver. Unlike full digital beamforming, the hybrid beamforming is separated into two different parts, one is the baseband digital precoding and the other is the analog beamforming implemented at RF domain \cite{YJiangYFeng_phase_2018,XYuJZhang_phase_2017,SSuhABasu_dft_2011}.
\\
 Therefore, beamforming needs information about the random wireless channel. Recently, many techniques have been studied to solve the channel estimation problem in mmWave systems by exploiting the sparsity of the mmWave channel. 
 \\
 The works \cite{Berraki_2014,Venugopal_2017,Shahmansoori_2018,Wu_2019} discuss channel parameter estimation based on compressed sensing (CS) theory by exploiting the sparsity, especially in mmWave channels. In \cite{Berraki_2014}, a CS-based channel parameter estimation is applied in a practical scenario and in a controlled environment, which turns out to be more attractive as compared to the exhaustive search methods. 
 	In \cite{Venugopal_2017}, the frequency selective case is considered and a CS-based algorithm is proposed to estimate the channel parameters assuming a hybrid architecture with quantized phase shifters.
  In \cite{Shahmansoori_2018}, a two-stage algorithm is introduced for position and orientation estimation. In the first stage, a modified version of the so-called distributed compressed sensing-simultaneous orthogonal matching pursuit (DCS-SOMP) algorithm \cite{Duarte_2005} is used on a predefined grid, while in the second stage the space-alternating generalized expectation maximization (SAGE) algorithm \cite{BHFleury_sage_1999} is used for the refinement of the channel parameter estimates. However, the paper does not shed light on the selection of beamformer, which is an important issue in the design of efficient hybrid architectures. In \cite{Wu_2019}, a low complexity orthogonal matching pursuit (OMP) algorithm is introduced by exploiting the correlation in the angular domain from  $\emph{a priori}$ statistical information about scattering. The algorithm, however, only considers flat fading channels.
 \\ 
 In \cite{D.Zhu_2016,D.Zhu_Globe_2016,D.Zhu_july_2017,D.Zhu_Dec_2017}, the auxiliary beam pair (ABP) method is proposed and discussed, which is based on a set of beam pairs to estimate the AoD. In the ABP method, by forming many beam pairs, a set of many ratios is calculated to estimate the corresponding AoD. The authors showed that the ABP method outperforms the grid of beams (GoB) method \cite{Singh_GoB_2015}. Furthermore, authors in \cite{Hu_2018} showed that the ABP method outperforms the standard CS techniques such as OMP \cite{Lee_2016} and adaptive codebook method \cite{Alkhateeb_2014}. 
Furthermore, in our previous work \cite{Fasim}, the DFT beams are implemented in the analog domain using a Butler matrix (BM) to estimate the channel via a maximum likelihood (ML) approach for a single path (frequency flat) scenario. \\
In this paper, we extend the previous work \cite{Fasim} and generalize it to the frequency selective channel model, for which the time-delay of each path is also taken into account in the algorithm design. This turns the channel parameter estimation into a non-linear optimization problem, which is solved with the SAGE algorithm as an approximation to the ML estimator in our scenario. In addition, we propose the column vectors of a DFT matrix, to probe the channel with constant amplitude zero autocorrelation (CAZAC) sequences \cite{BMPopovic_1992,Lindecazac_1992,Zadoff_1962}. Their constant amplitude property allows us to operate the power amplifier (PA) near to the saturation region and the set of beamforming vectors can be implemented at RF with the Butler matrix. This structure has an improved energy-efficiency and avoids the need for adaptive RF- phase shifters \cite{AGarcia_2016}. Our contributions can be summarized as follows:
\begin{itemize}
	\item We perform an ad-hoc estimation based on our proposed algorithm, herein referred to as  PREIDG (parameter estimation via interpolation based on a DFT grid) to coarsely estimate the model order (number of paths) and the parameters of these paths using a fixed look-up table (LUT). The proposed PREIDG algorithm outperforms the ABP method \cite{D.Zhu_2016,D.Zhu_Dec_2017}.
	\item The accuracy of the estimates can be further improved/refined if necessary, by using them as initialization of the SAGE algorithm to obtain the ML estimates. New expressions are derived for channel parameter estimation in the multiple-input-single-output (MISO) case.
	\item  We derive the Cram\'er-Rao lower bound (CRLB) to assess the performance of our two-stage estimation algorithm. 
\end{itemize}
\textbf{Notation:}
$a$ (lower case italic letters) denote scalars, $\mathbf{a}$ (bold lowercase letters) denote vectors and $\mathbf{A}$ (bold upper case letters) denote matrices;   $\B{I}_M$ represents identity matrix of size $M \times M$; $\B{1}_M$ denote all ones matrix of $M\times M$ dimensions; $\odot$ represents Hadamard product; $(\cdot)^\R{H}$ represents conjugate transpose; $\esp{\cdot}$ represents expectation; $\lfloor .\rfloor$ represents floor operator; $\R{tr}\{.\}$ represents trace of the matrix and $\textrm{mod}(.)$ represents modulus, respectively.
\section{System Model} \label{sec:model}
\subsection{System Architecture}
\begin{figure}[!t]
	\centering
	\includegraphics[width=0.7\linewidth]{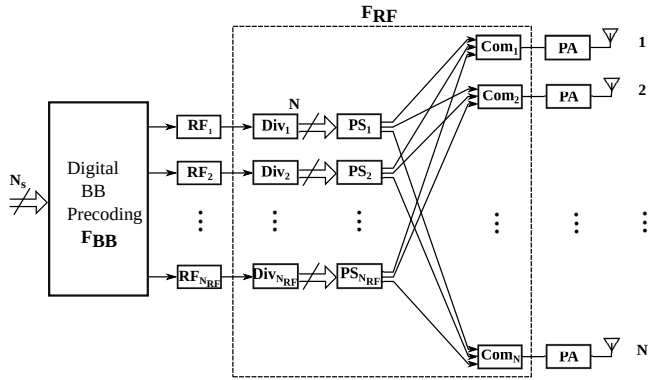}
	\caption{Fully connected analog phase shifting network.}
	\label{fig:a}
\end{figure}

\begin{figure}[!t]
	\centering
	\includegraphics[width=0.7\linewidth]{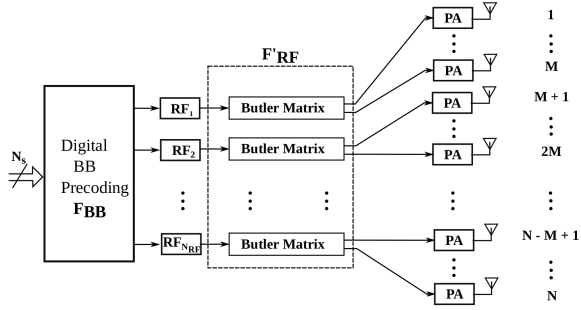}
	\caption{Partially connected analog phase shifting network.}
	\label{fig:b}
\end{figure}

\begin{figure}[!t]
	\centering
	\includegraphics[width=0.7\linewidth]{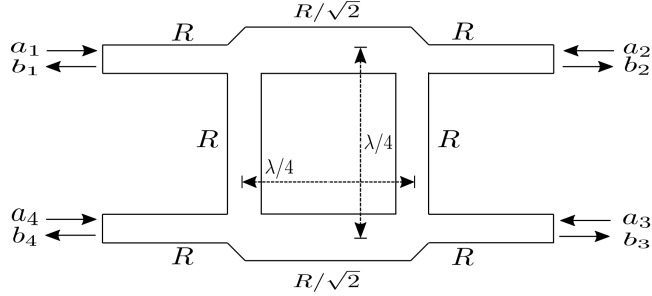}
	\caption{$90^\circ$ hybrid coupler.}
	\label{fig:aa}
\end{figure}

\begin{figure}[!t]
	\centering
	\includegraphics[width=1\linewidth]{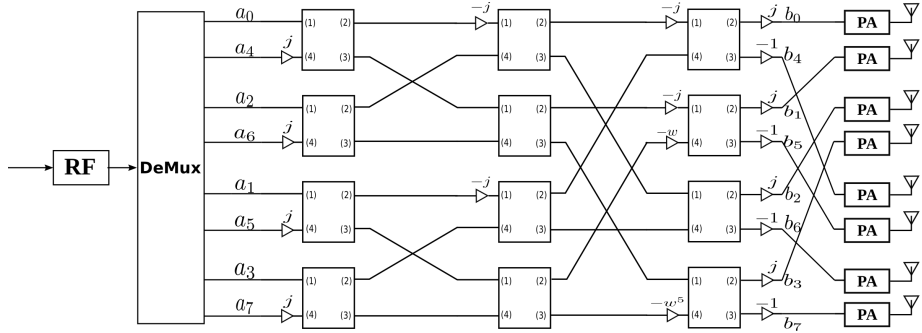}
	\caption{A $M\times M$, BM with $\frac{M}{2}\log_2M$, $90^\circ$ hybrids and fixed PSs only, where $w = \exp(-j\frac{2\pi}{M}) $ for $M=8$.}
	\label{fig:bb}
\end{figure}

To overcome the energy consumption and hardware complexity due to the use of massive MIMO and mmWaves, hybrid beamforming is introduced to cater for the reduction of RF-chains. In practice, there are two approaches in hybrid beamforming, one is based on a fully connected analog phase shifting (FCAPS) network and the other is based on a partially connected analog phase shifting (PCAPS) network. Both methodologies have their own pros and cons.
\\
In both the strategies dividers are used which divide the signals from the RF-chains to the many phase shifters (PSs). The combiner is only used in the FCAPS network. Although the power dividers are theoretically lossless, for instance, Wilkinson splitter, the combiners are not. 
\\
In the FCAPS network, $N_{RF} N$ adaptive phase shifters are used, where $N_{RF}$ is the total number of RF-chains and $N$ is the total number of transmit antennas as shown in \figurename{\;\ref{fig:a}}. In FCAPS network, at the input of each power amplifier and antenna, $N_{RF}$ signals have to be combined which introduces losses. The losses depend on the mutual correlation of the precoded signals to be combined which obviously depends on the estimated propagation channel.  On one hand the benefit of using FCAPS network, we can form narrow beams using all $N$ antennas simultaneously, but on the other hand, there are quite significant losses in the analog domain by the combination of signals which are channel dependent. For instance, if we have $N_{RF} = N_s = 2^q$, data streams, the combination of $N_s$ uncorrelated data streams will have a power loss of $q \times 3$ \si{\decibel}. On top of that, there come losses due to the parasitics especially at mmWaves. 
\\
To avoid these significant losses, we use the PCAPS approach as in \figurename{\;\ref{fig:b}}, where the combination of the signals in the RF-domain is avoided. In the PCAPS network, we divide the total number of transmit antennas $N$ into $N_{RF}$ sub-arrays, where each sub-array has $M$ antenna elements leading to $N = N_{RF} M$. In this approach, each sub-array gets its signal from one RF-chain. In this architecture, no combiner is required before the input of each antenna. With this, there will be no losses due to the combination of signals in the analog domain but also the number of adaptive phase shifters will be reduced from $N_{RF}  N$ to $N_{RF}  M$. 
\\
Furthermore, to avoid the implementation of adaptive phase shifters, the PCAPS network, can be implemented by introducing the BM, which is the analog implementation of the DFT matrix as shown in \figurename{\;\ref{fig:b}}.
\\
The $M$ beams of the DFT matrix are implemented using $\frac{M}{2}\log _2 M$ ($90^\circ$ hybrids) and a number of fixed phase shifters. The $90^\circ$ hybrid is theoretically a lossless 4 - port as shown in \figurename{\;\ref{fig:aa}}, which is described by the following scattering matrix 

\begin{equation}
\left[\begin{matrix}
b_1 \\
b_2 \\
b_3 \\
b_4
\end{matrix}\right] = 
\frac{1}{\sqrt{2}}
\left[\begin{matrix}
0 && -j && -1 && 0 \\
-j && 0 && 0 && -1 \\
-1 && 0 && 0 && -j \\
0 && -1 && -j && 0
\end{matrix}\right]
\left[\begin{matrix}
a_1 \\
a_2 \\
a_3 \\
a_4
\end{matrix}\right].   \label{sc1}
\end{equation}
Properly terminating all ports and with no incoming waves at ports 2 and 3 $(a_2 = 0 , a_3 = 0)$, resulting with no reflected waves at ports 1 and 4 $(b_1 = 0, b_4 = 0)$, therefore (1) can be reduced to 
\begin{equation}
\left[\begin{matrix}
b_2 \\b_3
\end{matrix}\right] = 
\frac{-1}{\sqrt{2}} 
\left[\begin{matrix}
j && 1 \\
1 && j
\end{matrix}\right]
\left[\begin{matrix}
a_1 \\
a_4
\end{matrix}\right].  \label{sc2}
\end{equation} 
An example of the BM structure with $M=8$ is shown in \figurename{~\ref{fig:bb}}, the output signals of the BM excited by one RF-chain connected to one input will be supplied to the $M$ power amplifiers and the $M$ antenna elements of each sub-array, which will result in utilizing one column of DFT matrix as beamforming vector.
\\
Now let us consider the input power $P_{in}$ to a $90^\circ$ hybrid at input port 1 and port 4 as $P_1$ and $P_4$ respectively. The $90^\circ$ hybrid will divide the power to the output port 2 and port 3, where only one of them will be used, as in \figurename{~\ref{fig:aa}}. The input power can be written as
\begin{equation}
P_{in} = P_1 + P_4  = \esp{|a_1|^2}+ \esp{|a_4|^2},
\end{equation}
and the output power at port 2 as 
\begin{equation}
P_{out_2} = \esp{|b_2|^2} = \frac{1}{2}\left(\esp{|a_1|^2} + \esp{|a_4|^2} - 2 \R{Im}\{\rho\} \sqrt{\esp{|a_1|^2} \esp{|a_4|^2}}\right),
\end{equation}
where $\rho$ is the correlation coefficient and is given as 
\begin{equation}
\rho = \frac{\esp{a_1a_4^\ast}}{\sqrt{\esp{|a_1|^2} \esp{|a_4|^2}}}.
\end{equation}
If both the signals at port 1 and port 4 are uncorrelated, i.e., $\rho = 0$, then $P_{out_2} = \frac{1}{2} P_{in}$, which means half of the power is lost due to the uncorrelated signals. However, in the BM architecture as shown in \figurename{\;\ref{fig:bb}}, always both the outputs of each $90^\circ$ hybrid are used and no power is lost.

\subsection{Signal Model}
We consider a single-user downlink scenario, where the user is equipped with a single antenna. The transmitter has PCAPS network architecture as shown in \figurename{~\ref{fig:b}} and equipped with a total number of $N$ antennas having $M$ number of antennas at each sub-array.
 The sub-arrays are controlled by $N_{\text{RF}}$ chains following $N = N_{\text{RF}} M$. The signal model based on the sub-array design and without loss of generality can be given as
\begin{equation}
\mathbf{y}_k^\R{T} = \sqrt{P_T}\,\sum_{r=1}^{R}\alpha_r\;\mathbf{a}^\R{H}(\mu_r)\mathbf{w}(\Phi_k)\mathbf{c}_k^\R{T}(\tau_r) + \mathbf{n}_k^\R{T} \in \mathbb{C}^{1 \times L}, \label{y}
\end{equation}
\eqref{y} is the received vector for the user based on one sub-array with the $k$th beamforming vector. $P_T$ represents the transmit power, $\alpha_r$ is the complex path coefficient of the given path $r$ and $\mathbf{n}$ is the random noise vector with Gaussian distribution $\mathbf{n} \sim \mathcal{C}\mathcal{N}(\bof{0}_{L \times 1},\sigma_n^2\B{I}_L)$, where $\sigma_n^2$ is the noise variance. $\B{c}_k(\tau_r)$ is the pilot sequence for $k$-th beamforming vector  $\B{w}(\Phi_{k})$, where $\tau_r$ is the delay of each path-$r$. The channel steering vector $\mathbf{a}(\mu_r)$  for uniform linear array (ULA) and beamforming vector $\mathbf{w}(\Phi_k)$ are given as
\begin{equation}
\mathbf{a}(\mu_r)=[1,e^{-j\mu_r},\dots,e^{-j(M-1)\mu_r}]^\R{T}\in\mathbb{C}^{M\times1},
\end{equation}
\begin{equation}
\mathbf{w}(\Phi_k) = \frac{1}{\sqrt{M}} \left[1, e^{-j\Phi_k}, \dots, e^{-j(M-1)\Phi_k}\right]^\R{T} \in \mathbb{C}^{M\times1},
\end{equation}
The $\mathbf{w}(\Phi_k)$ is $k+1$ th column of a DFT matrix of size $M$, where $\Phi_k = \frac{2\pi}{M}k, k=0,\dots,M-1$. $\mu_r$  is the spatial frequency of each path $r$ where $\mu_r = 2\pi\frac{d}{\lambda}\sin\theta_r $. $\theta_r$ is the angle of departure (AoD) of each path $r$ and $d$ is the distance between antenna elements respectively.
\\
The CAZAC sequence specific for each beamforming vector is $\mathbf{c}_k$, where each of the CAZAC sequence symbol is constructed as 
\begin{equation}
c(n) =  e^ {\left(j \frac{2\pi}{\sqrt{L}} \left(\textrm{mod} \left\{n,\sqrt{L}\right\} + 1\right)\left(\left\lfloor \frac{n}{\sqrt{L}}\right\rfloor+1\right)+j\frac{\pi}{4}\right)},
\end{equation} 
where $n\in\left\{0,1,\dots,L-1\right\}$ and  $\B{c}_k(n) = \B{c}(n-k)$.
By considering the length of the CAZAC sequence $L=16$ helps in forming the CAZAC symbols as QPSK symbols $c(n) \in \left\{\frac{1}{\sqrt{2}}\left(\pm 1\pm j\right)\right\}$, which is useful because of the constant modulus allow us to operate the PA near the saturation region. The CAZAC sequence for first beamforming vector is represented as 
\begin{equation}
\mathbf{c}_0(0) = \left[c(0),c(1),\dots,c(L-1)\right]^\R{T} \in \mathbb{C}^{L \times 1}.
\end{equation}
 $\B{c}_k$, where $k = 0,\dots,M-1$ are shifted wrap around versions of $\B{c}_0$ and assigned each wrap around to a specific beamforming vector. 
\\
By probing the channel with each beamforming vector $\mathbf{w}(\Phi_k)$ with its corresponding CAZAC sequence $\mathbf{c}_k$ and then collecting all the received vectors $\mathbf{y}_k$ in matrix $\mathbf{Y}$ can be represented as 

\begin{equation}
\mathbf{Y} = \left[\begin{array}{c}
\B{y}_0^\R{T} \\
\B{y}_1^\R{T} \\
\vdots \\
\B{y}_{M-1}^\R{T}
\end{array}\right] = \nonumber
\end{equation}
\begin{equation}
=\sqrt{P_T} \sum_{r=1}^{R}\alpha_r \B{A}(\mu_r)\underbrace{\left[ \begin{array}{c}   \B{c}_0^{\R{T}}(\tau_r) \\  \B{c}_1^{\R{T}}(\tau_r) \\ \vdots  \\ \B{c}_{M-1}^{\R{T}}(\tau_r)  \end{array} \right]}_{=\B{C}(\tau_r)} + \underbrace{\left[ \begin{array}{c} \B{n}_0^{\R{T}}  \\ \B{n}_1^{\R{T}} \\ \vdots \\ \B{n}_{M-1}^{\R{T}} \end{array} \right]}_{=\B{N}} \in \mathbb{C}^{M \times L}, \label{Y}
\end{equation}
where $\mathbf{A}(\mu_r) = \text{diag}\{\mathbf{a}^\R{H}(\mu_r)\mathbf{w}(\Phi_k)\}_{k=0}^{M-1}$ and the noise covariance matrix is  $\B{R} = \esp{ \R{vec}\{\B{N}\} \; \R{vec}\{\B{N}\}^{\R{H}} }=\sigma_n^2 \B{I}_{ML}$. 

\section{Parameter Estimation via Interpolation based on a DFT Grid (PREIDG) based coarse estimation}
\begin{figure}[!t]
	\centering
	\includegraphics[width=0.4\linewidth]{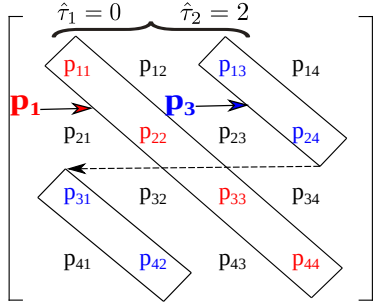}
	\caption{An example of integer delay estimation $\hat{\tau_i}$ using power matrix $\B{P}$ as in \eqref{Pmatrix} where $\B{p}_1$ and $\B{p}_3$ are the  two diagonals, for which \eqref{pvector} is fulfilled.}
	\label{fig:powertable}
\end{figure}

\begin{figure}[!t]
	\centering
	\includegraphics[width=0.7\linewidth]{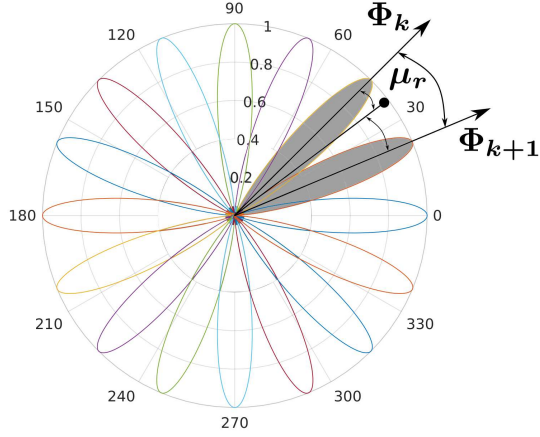}
	\caption{An example of AoD estimation using a DFT matrix based beamforming via PREIDG.}
	\label{fig:PTDFT}
\end{figure}

We are now probing the channel with all $M$ beamforming vectors, one at a time with one specific CAZAC sequence of length $L$, preferably $L=M$. There is a strict correspondence between $\B{w}(\Phi_{k})$ and $\B{c}_k$. This way the user equipment (UE) observes $M$ consecutive receive sequences $\B{y}_k$ \eqref{y} and multiplies each of them with the already stored CAZAC sequence $\B{c}_k^\ast$. This can be cast in a matrix 
\begin{equation}
\B{Z} = \B{Y}\B{C}^\R{H}(0)=\sqrt{P_T}\sum_{r=1}^{R}\alpha_r\B{A}(\mu_r)\B{C}(\tau_r)\B{C}^\R{H}(0) + \B{N}\B{C}^\R{H}(0), \label{Zmatrix}
\end{equation} 
For simplicity, let us first assume that each of the $R$ AoD's, $\mu_r$ is equal to one of the $\Phi_k$. Then we get
\begin{align}
\B{A}(\mu_r) &= \text{diag}\left\{\B{a}^\R{H}(\mu_r) \B{w}(\Phi_k)\right\}_{k=0}^{M-1}\Bigg |_{\mu_r = \Phi_{k_r}} \nonumber \\
&=\sqrt{M}\,\B{e}_{k_r+1}\B{e}_{k_r+1}^\R{T},
\end{align}
where $\B{e}_{k_r+1}$ is the $M$- dimensional $(k_r +1)$ canonic unit vector. In addition, let us assume that each of the $R$ delays are integer multiples of the symbol period leading to
\begin{equation}
\B{C}(\tau_r)\B{C}^\R{H}(0) \,\,\bigg |_{\tau_r = i_r} = M \B{P}_{i_r},
\end{equation} 
where $\B{P}_{i_r}$ is the $M \times M$, permutation matrix of the following form, by using the canonic unit vector $\B{e}$,
\begin{align}
\B{P}_{i_r} &= \sum_{j=1}^{M}\B{e}_j \B{e}^\R{T}_{j+i_r} \nonumber \\
&=\left[ \begin{matrix}
0 & \dots & 0 & 1 & 0 & 0 & \dots & 0 \\
0 & \dots & 0 & 0 & 1 & 0 & \dots & 0 \\
\vdots & \dots & \vdots & \vdots & \dots & \ddots & \dots & \vdots \\ 
\vdots & \dots & \vdots & \vdots & \dots & \dots & \ddots & \vdots \\
0 & \dots & \vdots & \vdots & \dots & \dots & \dots & 1 \\
1 & \dots & \vdots & \vdots & \dots & \dots & \dots & 0 \\
\vdots & \ddots & \vdots & \vdots & \dots & \dots & \dots & \vdots \\
0 & \dots & 1 & \dots & \dots & \dots & \dots & 0 \\
\end{matrix}
\right],
\end{align}
with $\B{P}_0 = M \B{I}_M$. In this simplified case, the matrix $\B{Z}$ is written as
\begin{align}
\B{Z} &= \sqrt{P_T}M\sqrt{M}\sum_{r=1}^{R} \alpha_r\left(\B{e}_{k_r +1}\B{e}_{k_r +1}^\R{T}\sum_{j=1}^{M}\B{e}_j\B{e}^\R{T}_{j+i_r}\right) + \B{N}\B{C}^\R{H}(0) \nonumber \\
&= \sqrt{P_T}M\sqrt{M}\sum_{r=1}^{R}\alpha_r\B{S}_r + \B{N}\B{C}^\R{H}(0).
\end{align} 
where $\B{S}_r$ is a matrix, where only the entry in the $(k_r +1)$ row and $\mod(k_r + i_r +1,M)$ column is equal to one, while all the other entries are zero. The integers $k_r$ and $i_r$ identify the AoD and $\tau$ of the $r$th multipath component. 
\\
Finally, we compute the post correlation power matrix for the general form as
\begin{align}
\B{P} = \esp{\B{Z}\odot\B{Z}^\ast} &= P_T M^3 \sum_{r=1}^{R}|\alpha_r|^2 \B{S}_r \odot \B{S}_r^\ast + \sigma_n^2 M\B{1}_M. \label{Pmatrix}
\end{align}
The post correlation receive signal to noise ratio (SNR) is enhanced by a process gain $M$ from correlation and by an antenna gain of $M$ from the array, which is valid for our simplistic assumption, that both $\mu_r$ and $\tau_r$ are on the grid of the beamforming vectors as well as of the symbol timing. In any realistic scenario, this will not be true and the matrices $\B{S}_r$ will not be strictly sparse with only one non-zero entry. But the power matrix, which is available at the user, will still provide useful information about the model order, i.e. the number of the paths/wavefronts and the parameters $\sqrt{P_T}\alpha_r,\mu_r,\tau_r$ of each path. We exploit the power matrix by searching the (wrap-around) diagonals $\B{p}_i,i=1,\dots,M$ of $\B{P}$ for the largest entry as shown in \figurename{\;\ref{fig:powertable}} for $M=4$.
\\
We have
\begin{equation}
\B{p}_i^\R{T}=\left[p_{1,i}\, ,\,p_{2,\textrm{mod}(i,M)+1},\dots,p_{M,\textrm{mod}(i+M-2,M)+1}\right]^\R{T} \in \mathbb{R}^M, \label{diagonal_P}
\end{equation}
where $i = 1,\dots,M$, with $\B{p}_1$ being the main diagonal of $\B{P}$. For every diagonal $\B{p}_i$ we check, whether the largest element is above a certain threshold $G$,
\begin{equation}
\max_{k=1,\dots,M}\left(p_{k,\textrm{mod}(i+(k-2),M)+1}\right)\ge G, \,\, i=1,\dots,M. \label{pvector}
\end{equation}
\\
$G$ should make sure that we have a signal above the noise floor, which is $\sigma_n^2 M$ \eqref{Pmatrix}. The integer delay $\tau_{i_r}$ for each path $r$ can be found as shown in \figurename{\;\ref{fig:powertable}},
\begin{equation}
\hat{\tau}_{i_r} =  i_r-1. \label{integerdelay}
\end{equation}
 For the main diagonal $\B{p}_1$, this will always be fulfilled assuming a LOS path, which has relative delay $\tau_i = i_1-1\big |_{i_1=1} =0$. The number of diagonals $\B{p}_i$, which fulfill \eqref{pvector} is the model order $\hat{R}$ and we have a coarse estimate $\hat{\tau}_{i_r}$ for each path, where we drop index $r$ for simplicity.
\\
Since the real spatial frequency of each detected path will be somewhat in between two spatial frequencies $\Phi_k < \mu_r <\Phi_{k+1}$ as in \figurename{\;\ref{fig:PTDFT}}, there will be two significant adjacent entries along the diagonal, which we denote $P_k$ and $P_{k+1}$.
\\
\textbf{LUT:}
 We get an estimate of $\mu_r$, by interpolation with $K+1$ spatial frequencies $\mu_l$, generated as follows
\begin{align}
\mu_l &= \Phi_k + l\Delta_{\mu}, \,\,\, l=0\dots,K, \\
\Delta_\mu &= \frac{\Phi_{k+1}-\Phi_k}{K} = \frac{2\pi}{MK}, 
\end{align}
and compute the hypothetical noise free normalized power for those angles, $\mu_l$ 
\begin{align}
P_{k,l}&=\left| \mathbf{a}^\text{H}(\mu_l) \mathbf{w}(\Phi_k)\right|^2, \\
P_{k+1,l}&=| \mathbf{a}^\text{H}(\mu_l) \mathbf{w}(\Phi_{k+1})|^2, 
\end{align}
with ratios
\begin{equation}
\Delta_{l} = \sqrt{\frac{P_{k,l}}{P_{k+1,l}}}. \label{diffinterpol}
\end{equation}
Since $P_{k,l}$ and $P_{k+1,l}$ are independent of $k$, we need only $K+1$ ratios to provide $\Delta_l$ in a LUT. Note that the proposed LUT is different from the idea discussed in \cite{Candes_2008}, which is based on a re-weighted \textit{l}-1 minimization problem. The entries of the measurement matrix are independent and identically distributed (i.i.d) Gaussian random variables, and besides the observed data is contaminated by noise.  This approach is significantly different from what we have considered in the generation of the fixed LUT.
\\
\textbf{\textit{~Coarse estimation of $\hat{\mu}_r$:}}
Now we are selecting those indices $l$ and $l+1$, where the corresponding ratios $\Delta_{ {l}}$ and $\Delta_{ {l+1}}$ are closest to the ratio \eqref{diffpwr}
\begin{equation}
\Delta= \sqrt{\frac{P_k}{P_{k+1}}},
\label{diffpwr}
\end{equation}	
and get the estimated spatial frequencies 
\begin{equation}
\hat{\mu}_r= \mu_l + b\Delta_\mu = \Phi_k + \Delta_\mu(l+b),
\label{muhat}
\end{equation}
with 
\begin{equation}
b = \frac{\Delta_{{l}}-\Delta}{\Delta_{{l}}-{\Delta_{{l+1}}}}.
\label{interpol}
\end{equation}
Finally, \eqref{muhat} can be converted to an estimated azimuth angle in degrees
\begin{equation}
\hat{\theta}_r= \begin{cases}
\arcsin(\frac{\hat{\mu}_r}{\pi}),\; 0\le \hat{\mu}_r\le \pi, \\
\arcsin(\frac{\hat{\mu}_r-2\pi}{\pi}),\; \pi < \hat{\mu}_r\le 2\pi.
\end{cases} \label{thetahat}
\end{equation} 
The interpolation between $\Phi_k$ and $\Phi_{k+1}$ can go wrong in some cases especially if the signal level is weak and $\mu_r$ is close to letting say $\Phi_k$. In such a case $P_k$ may be quite large, but $P_{k+1}$ may be close to the noise floor. Therefore it is not clear whether $P_{k+1}$ or rather $P_{k-1}$ is the second largest power from the signal, which is masked by the noise. Therefore we check whether 
\begin{equation}
|P_{k +1} - P_{k-1}| \le \frac{\sigma_n^2}{v}. \label{thresholdratio}
\end{equation}	
If \eqref{thresholdratio} is fulfilled, then it is not worthwhile to interpolate at all, but simply choose $\hat{\mu}_r = \Phi_k$. The value of $v = 3$, has been heuristically chosen based on numerical experiments. Keep in mind, that the LUT once generated is fixed for every spatial frequency $\mu_r$. 
\\
After having estimated all AoDs, the model order estimation may be refined, because of the integer estimation of the delays. One non-integer delay may have lead to two adjacent integer delays, and both of them will have the same AoD estimate. If this occurs, we drop one of the two delays.
\\
Now we can put together the whole coarse estimation approach for AoD estimation in Algorithm~\ref{Algortithm}. The only thing the UE has to feedback to the BS is $\Delta$ in \eqref{diffpwr} which is one real number, and one index with $\log_2 M$ bits for each path $r$. The rest of the computation can be done by BS for each AoD.
\begin{algorithm}
	\caption{Proposed coarse estimation based on the PREIDG algorithm.}
	\label{Algortithm}
	\begin{algorithmic}[1]
		\Require $\B{Y}$  \eqref{Y}
		\State The UE received $\B{Y}$ and get $\B{p}_i$ for each path $r$ \eqref{Pmatrix} 				
		\State Calculate $\Delta$ as in \eqref{diffpwr}
		\State Find $l$ such that $\Delta_{{l}} \ge \Delta \ge \Delta_{{l+1}}$
		\State Calculate constant $b$ as in \eqref{interpol}					
		\State \Return  $\hat{\mu}_r$ and $\hat{\theta}_r$ as in \eqref{muhat} and \eqref{thetahat}.
	\end{algorithmic}
\end{algorithm}

\begin{figure}[!t]  
	\centering
	\includegraphics[width=0.6\linewidth]{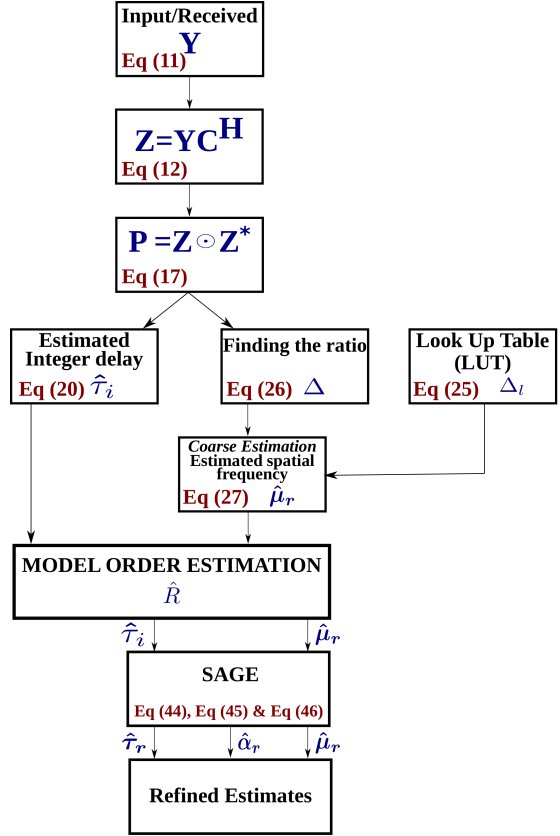}
	\caption{A Flowchart for the two-stage estimation algorithm.}
	\label{fig:flowchart}
\end{figure}

\section{Maximum Likelihood Estimation using SAGE}
The channel parameters obtained using coarse estimation based on the PREIDG algorithm can be further improved using an iterative procedure which is initialized by the ad-hoc estimates found by PREIDG. This is a multidimensional, non-linear optimization problem \eqref{nonlinearproblem}, which can be solved by space-alternating generalized expectation maximization (SAGE) algorithm. We use the standard formulation of the SAGE algorithm \cite{BHFleury_sage_1999} to solve and derive new expressions for our non-linear problem considering important hardware constraints.
\\
 To make sure that the SAGE algorithm converges to the global optimum with low computational complexity. Initializing SAGE with the coarse estimation is indispensable. Furthermore, SAGE will also help us to estimate the non-integer delay, unlike the coarse estimation in addition to the more refined complex path gain for each path $r$. 
\\
We treat the observed data $\B{Y}$ as a random variable, parametrized by a Gaussian probability density function (pdf) with an unknown channel parameter vector $\bof{\eta}$ as
\begin{equation}
\bof{\eta} = \left[\sqrt{P_T}\R{Re}\{\bof{\alpha}\}^\R{T},\sqrt{P_T}\R{Im}\{\bof{\alpha}\}^\R{T},\bof{\mu}^\R{T},\bof{\tau}^\R{T}\right]^\R{T}, \label{Gparametervector}
\end{equation}
where 
\begin{equation}
\sqrt{P_T}\R{Re}\{\bof{\alpha}\} = \left[\sqrt{P_T}\R{Re}\{\alpha_1\},\dots,\sqrt{P_T}\R{Re}\{\alpha_r\},\dots,\sqrt{P_T}\R{Re}\{\alpha_R\}\right]^\R{T},
\end{equation}
\begin{equation}
\sqrt{P_T}\R{Im}\{\bof{\alpha}\} = \left[\sqrt{P_T}\R{Im}\{\alpha_1\},\dots,\sqrt{P_T}\R{Im}\{\alpha_r\},\dots,\sqrt{P_T}\R{Im}\{\alpha_R\}\right]^\R{T},
\end{equation}
\begin{equation}
\bof{\mu} = \left[\mu_1,\dots,\mu_r,\dots,\mu_R\right]^\R{T},
\end{equation}
\begin{equation}
\bof{\tau} = \left[\tau_1,\dots,\tau_r,\dots,\tau_R\right]^\R{T},
\end{equation}
with the likelihood function given as 
\begin{align}
\R{L}(\B{Y};\bof{\eta}) = \frac{1}{\pi^{ML} \det\B{R} }\; \exp \left(-\R{vec}\left\{\B{Y} - \sqrt{P_T}\sum_{r=1}^{R}\alpha_r \B{A}(\mu_r)\B{C}(\tau_r)\right\}^\R{H} \right. \nonumber \\
\left. \B{R}^{-1}\;\R{vec}\left\{\B{Y} - \sqrt{P_T}\sum_{r=1}^{R}\alpha_r \B{A}(\mu_r)\B{C}(\tau_r)\right\}\right). \label{nonlinearproblem}
\end{align}
SAGE uses the observable but incomplete data space mentioned in \eqref{Y} to estimate the parameters of the superimposed $R$ wavefronts.
\begin{equation}
\B{Y} = \bof{f}(\B{X})=\bof{f}\left(\left[\B{X}_1,\dots,\B{X}_R\right]\right) = \sum_{r=1}^{R}\B{X}_r, 
\end{equation}
where $\B{X}$ is the complete but unobservable data space. $\B{X}_r$ is called a hidden data space.
\\
In each iteration the following expectation and maximization steps are performed. The $\hat{\tau}_i$ and $\hat{\mu}_r$ are the ad-hoc estimates obtained using \eqref{integerdelay} and \eqref{muhat} which will be used to initialize SAGE, while $\hat{\alpha}_r = 0$ is assumed.
\\
\textbf{\textit{~Expectation Step:}}
The conditional expectation of the hidden data space can be calculated based on the incomplete data space $\B{Y}$ and the previous estimation $\bof{\hat{\eta}}$.
\begin{equation}
\hat{\B{X}}_r = \mathbb{E}_{\B{X}_r}[\B{X}_r|\B{Y};\hat{\bof{\eta}}]= (1-\beta_r)\B{S}_r(\hat{\bof{\eta}}_r) + \beta_r \left(\B{Y} - \sum\limits_{\substack{r^\prime =1 \\ r^\prime \neq r}}^R\B{S}_{r^\prime}(\hat{\bof{\eta}}_{r^\prime})\right),
\end{equation}
where
\begin{equation}
\B{S}_r(\bof{\eta}_r) = \sqrt{P_T}\alpha_r\B{A}(\mu_r)\B{C}(\tau_r),
\end{equation}
and
\begin{equation}
\bof{\eta}_r = \left[\sqrt{P_T}\R{Re}\{\alpha_r\},\sqrt{P_T}\R{Im}\{\alpha_r\},\mu_r,\tau_r\right],
\end{equation}
and $\beta_r$ controls the convergence rate. Assuming $\beta_r = 1$ the estimated hidden data space is estimated as
\begin{equation}
\B{\hat{X}}_r = \B{Y} - \sum\limits_{\substack{r^\prime =1 \\ r^\prime \neq r}}^R \B{S}_{r^\prime}(\bof{\hat{\eta}}_{r^\prime}).
\end{equation}
\\
\textbf{\textit{~Maximization Step:}}
To compute a refined $\hat{\bof{\eta}}_r$, the following optimization problem has to be solved,
\begin{equation}
\hat{\bof{\eta}}_r = \arg\max_{\bof{\eta}_r}\mathbb{E}_{\B{X}_r}\left[\ell (\B{X}_r;\bof{\eta}_r)|\B{Y};\hat{\bof{\eta}}\right], \label{costfunction}
\end{equation}
where 
\begin{equation}
\ell (\B{X}_r;\bof{\eta}_r) = \ln \,(\R{L}\,(\B{X}_r;\bof{\eta}_r)),
\end{equation}
is the log-likelihood function. This non-linear optimization problem \eqref{costfunction} can be solved iteratively and sequentially. The new derivation of the below equations are given in \ref{Derivation_ML}.
\\
The delay estimation $\hat{\tau}_r$ can be iteratively maximized as
\begin{equation}
\hat{\tau}_r= \arg \max_{\tau_r} \left\{\frac{ \left|\R{tr}\left\{\B{C}^\R{H}(\tau_r)\B{A}^\R{H}(\hat{\mu}_r)\hat{\B{X}}_r\right\}\right|^2  }{\beta_r\sigma_n^2 \; \R{tr}\{\B{C}^\R{H}(\tau_r)\B{A}^\R{H}(\hat{\mu}_r)\B{A}(\hat{\mu}_r)\B{C}(\tau_r)\}}\right\}, \label{tauestimator}
\end{equation}
similarly the spatial frequency $\hat{\mu}_r$ can be iteratively maximized as
\begin{equation}
\hat{\mu}_r= \arg \max_{\mu_r} \left\{\frac{ \left|\R{tr}\left\{\B{C}^\R{H}(\hat{\tau}_r)\B{A}^\R{H}(\mu_r)\hat{\B{X}}_r\right\}\right|^2  }{\beta_r\sigma_n^2 \; \R{tr}\{\B{C}^\R{H}(\hat{\tau}_r)\B{A}^\R{H}(\mu_r)\B{A}(\mu_r)\B{C}(\hat{\tau}_r)\}}\right\}, \label{muestimator}
\end{equation}
while in the end $\hat{\sqrt{P_T}\alpha}_r$ can be analytically found as
\begin{equation}
\hat{\sqrt{P_T}\alpha}_r = \frac{\R{tr}\left\{\B{C}^\R{H}(\hat{\tau}_r)\B{A}^\R{H}(\hat{\mu}_r)\hat{\B{X}}_r\right\}}{ \R{tr}\{\B{C}^\R{H}(\hat{\tau}_r)\B{A}^\R{H}(\hat{\mu}_r)\B{A}(\hat{\mu}_r)\B{C}(\hat{\tau}_r)\}}. \label{alphaestimator}
\end{equation}
One iteration of the SAGE algorithm is defined as a full update of all parameters of the parameter vector $\bof{\eta}$. To make sure that SAGE converges to the global optimum, it is important to initialize the SAGE algorithm with a good coarse estimation. We initialize SAGE with the coarse estimated spatial frequency $\hat{\mu}_r$ and integer delay $\hat{\tau}_i$ as shown in \figurename{\;\ref{fig:flowchart}}.
\\
The stopping thresholds for convergence of SAGE are defined as
\begin{align}
\R{T}_1 &= \frac{|\hat{\mu}_{r_p} - \hat{\mu}_r|}{|\hat{\mu}_r|}, \\
\R{T}_2 &= \frac{|\hat{\tau}_{r_p} - \hat{\tau}_r|}{|\hat{\tau}_r|}, \\
\R{T}_3 &= \frac{|\hat{\sqrt{P_T}\alpha}_{r_p} - \hat{\sqrt{P_T}\alpha}_r|}{|\hat{\sqrt{P_T}\alpha}_r|},
\end{align}
 where $\hat{\mu}_{r_p},\hat{\tau}_{r_p},\hat{\sqrt{P_T}\alpha}_{r_p}$, are the previous estimates of spatial frequency, delay time and complex path gain. The stopping criteria for SAGE is satisfied, when \eqref{thershold} is fulfilled
 \begin{equation}
 \max \left\{\R{T}_1,\R{T}_2,\R{T}_3 \right\} \le \Gamma, \label{thershold}
 \end{equation}
 where $\Gamma$ is the stopping threshold.
   The coarse parameters are fed to the SAGE algorithm after refinement of the model order estimation as shown in \figurename{\;\ref{fig:flowchart}}.
\\
All these computations will be done by the UE and for each path $r$ four real numbers i.e. $\hat{\mu}_r,\hat{\sqrt{P_T}\alpha}_r, \hat{\tau}_r$ will be quantized and feedback to the BS for further hybrid beamforming. Of course, the single antenna UE has to have enough power to feedback this information to the BS. The flowchart for the two-stage algorithm is shown in \figurename{\;\ref{fig:flowchart}}.

\section{Derivation of the CRLB}
In this section, we derive the CRLB for spatial frequency ($\mu_r$), complex path gain ($\sqrt{P_T}\alpha_r$) and delay ($\tau_r$) estimation. 
\\
Assuming $\bof{\hat{\eta}}$ as an unbiased estimate of $\bof{\eta}$, then the variance, $\R{var}$ of the estimation error for the different parameters can be lower-bounded by the diagonal elements of the inverse of Fisher information matrix (FIM) represented as $\B{F}(\bof{\eta})$ \cite{SMKay}
\begin{equation}
\R{var}(\hat{\eta}_i)  \ge \left[\B{F}^{-1}(\bof{\eta})\right]_{i\,i}.
\end{equation}
The bound on the error is calculated as 
\begin{equation}
\sqrt{\text{CRLB}(\hat{\eta}_i)} = \sqrt{\left[\B{F}^{-1}(\bof{\eta})\right]_{i\,i}}. \label{crlb}
\end{equation}
The FIM for complex data is given as \cite{SMKay},
\begin{align}
\left[\B{F}(\bof{\eta})\right]_{ij} =
\frac{2}{\sigma_n^2}\,\, \R{Re}\,\,\left\{\R{tr}\left\{\frac{\partial \B{S}^\R{H}(\bof{\eta})}{\partial \eta_{i}}\,\,\frac{\partial\B{S}(\bof{\eta})}{\partial\eta_{j}}\right\}\right\}, \label{complexdata}
\end{align}
with $\B{S}(\bof{\eta})$ is defined as
\begin{equation}
\B{S(\bof{\eta})} = \sum_{r=1}^{R}\B{S}(\bof{\eta}_r) = \sqrt{P_T}\;\sum_{r=1}^{R}\alpha_r\,\B{A}(\mu_r)\,\B{C}(\tau_r).
\end{equation} 
The FIM can be structured as
\begin{align}
&\B{F}(\bof{\eta}) = \nonumber \\
&\left[\begin{matrix}
\B{F}_{\R{Re}\left\{\bof{\alpha}\right\}\R{Re}\{\bof{\alpha}\}} && \B{F}_{\R{Re}\{\bof{\alpha}\}\R{Im}\{\bof{\alpha}\}} && \B{F}_{\R{Re}\{\bof{\alpha}\}\bof{\mu}} && \B{F}_{\R{Re}\{\bof{\alpha}\}\bof{\tau}} \\
\B{F}^\R{T}_{\R{Re}\{\bof{\alpha}\}\R{Im}\{\bof{\alpha}\}} && \B{F}_{\R{Im}\{\bof{\alpha}\}\R{Im}\{\bof{\alpha}\}} && \B{F}_{\R{Im}\{\bof{\alpha}\}\bof{\mu}} && \B{F}_{\R{Im}\{\bof{\alpha}\}\bof{\tau}} \\
\B{F}^\R{T}_{\R{Re}\{\bof{\alpha}\}\bof{\mu}} && \B{F}^\R{T}_{\R{Im}\{\bof{\alpha}\}\bof{\mu}} && \B{F}_{\bof{\mu\mu}} && \B{F}_{\bof{\mu\tau}} \\
\B{F}^\R{T}_{\R{Re}\{\bof{\alpha}\}\bof{\tau}} && \B{F}^\R{T}_{\R{Im}\{\bof{\alpha}\}\bof{\tau}} && \B{F}^\R{T}_{\bof{\mu\tau}} && \B{F}_{\bof{\tau\tau}} 
\end{matrix}\right]. \label{FIM}
\end{align}
The block matrices of $\B{F}(\bof{\eta})$ are derived in \ref{Derivation_CRLB}.

\section{Numerical Results}
In this section, the performance of the proposed two-stage algorithm is evaluated, assessed with CRLB and compared with ABP. The transmitter is deployed with a subarray antenna structure with $M = 16$ and a total number of antennas $N =N_{\text{RF}}M$ as a ULA with $\lambda/2$ inter-element spacing.\\
We assume the bandwidth of a system, $B = \SI{200}{\mega\hertz}$, from which one symbol time can be calculated as $T_s = \SI{5}{\nano\second}$. The system is operating at a carrier frequency of $f_c = \SI{28}{\giga\hertz}$. 32 pilot symbols are used per beamforming vector which ends up with a total number of 512 pilot symbols.
The distance for LOS  is uniformly distributed as \cite{SamimiRappaport_2016} 
\begin{equation}
D_{\text{los}} \sim \mathrm{U}(\SI{30}{ \meter},\SI{60}{\meter}),
\end{equation}
while the NLOS distances are distributed as 
\begin{equation}
D_{\text{nlos}} = D_{\text{los}} + \Delta_{\text{nlos}},
\end{equation}
where the $\Delta_{\text{nlos}}$ is the difference of NLOS and LOS paths and can be distributed as $\Delta_{\text{nlos}}\sim \R{U}(\SI{4.5}{\meter},\SI{24}{\meter})$,  which gives the delay between 3 and 16 symbols. The length of CAZAC sequence restricts the maximum delay difference estimated.
The path loss ($PL$) is calculated as \cite{SamimiRappaport_2016}
\begin{equation}
PL\text{(\si{\decibel})} = 10  \bar{n} \log _{10}\left(\frac{D}{D_0}\right), \label{Pathloss}
\end{equation}
where $\bar{n}$ is the path loss exponent which is  assumed $2.1$ for LOS and $2.4$ for NLOS paths and $D_0 = \SI{1}{\meter}$. 
The complex path gain for LOS is assumed such that $\alpha_1 = 1$. To
calculate the complex path gains for NLOS, we need to use the path loss \eqref{Pathloss} for LOS ($PL_{\text{los}}$) and NLOS ($PL_{\text{nlos}}$), 
\begin{equation}
\gamma_r = \sqrt{\frac{PL_{\text{los}}}{PL_{\text{nlos}}} },\label{PLratio}
\end{equation}
where $\gamma_r$ is the ratio of LOS and NLOS paths and is the magnitude of the complex path coefficient. The complex path gain, $\alpha_r$ for each path $r$ can be obtained as
\begin{equation}
\frac{\alpha_r}{\alpha_1} = \gamma_r e^{j\phi_r},
\end{equation} 
where $\phi_r$ is the phase of the complex coefficient of path $r$ and is uniformly distributed as $\phi_r \sim \mathrm{U}(0,2\pi)$ \cite{SamimiRappaport_2016}. $\alpha_1$ is assumed as 1, because what matters is the ratio of the LOS path as compared to the NLOS path.
\\
The AoD  for both LOS path and NLOS paths are generated from a uniform distribution, i.e. $\theta_r^\circ \sim \mathrm{U}(-60^\circ , +60^\circ)$ (one sector of a cell)  as given in \cite{SamimiRappaport_2016}. The noise variance is kept as $\sigma_n^2 = 1$.
\\
The SNR is defined as
\begin{equation}
\text{SNR} = \frac{P_T |\alpha_1|^2}{\sigma_n^2}, 
\end{equation}
The root mean square error $(\R{RMSE})$ for $\theta_r$, $\sqrt{P_T}\alpha_r$ and $\tau_r$ are calculated as
\begin{align}
\R{RMSE}(\hat{\theta}_r) &= \sqrt{\esp{\vert\theta_r - \hat{\theta}_r\vert^2}}, \\
\R{RMSE}(\hat{\sqrt{P_T}\alpha_r}) &= \sqrt{\esp{\left\vert\frac{\sqrt{P_T}\alpha_r - \hat{\sqrt{P_T}\alpha_r}}{\sqrt{P_T}\alpha_r}\right\vert^2}}, \\
\R{RMSE}(\hat{\tau}_r) &=\sqrt{\esp{\vert\tau_r - \hat{\tau}_r\vert^2}}. 
\end{align}
\begin{figure}[!t]
	\centering
	\includegraphics[width=0.7\linewidth]{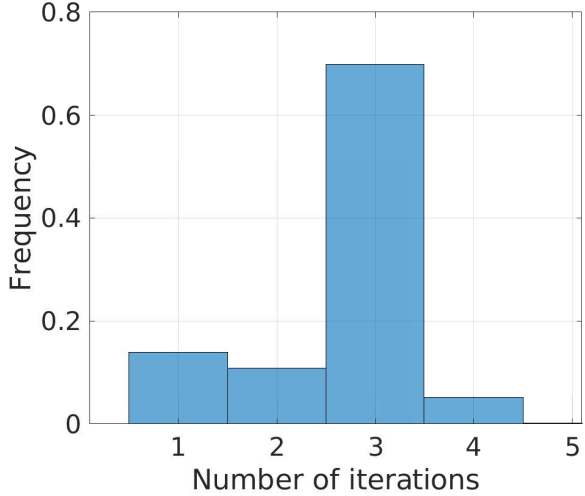}
	\caption{Histogram for the number of iterations of SAGE.}
	\label{fig:histogram}
\end{figure} 
In calculating RMSE, we are not calculating true expected values, but we used numerical averaging over 10000 realizations. 
\\
We are calculating CRLB for all realizations as given in \eqref{crlb} and then obtain a numerical average over 10000 channel scenarios. Finally, we take the square root to compare with RMSE respectively. The length of the LUT is chosen as $K = 101$. 
\\
 In the ABP approach, the formation of auxiliary pairs of beams is important to estimate the corresponding AoD. For this, a specific spacing difference between the beams named as $\delta$ is important and need to be kept constant as in \cite{D.Zhu_2016,D.Zhu_Dec_2017}. The spacing is given as, $\delta =  \frac{2m\pi}{M}$, where $m$ can be chosen from the set as $m = 1,\dots, \frac{M}{4}$. We choose $m=1$ and get $\delta = \frac{\pi}{8}$.  To keep $\delta = \frac{\pi}{8}$ constant in all the beam pairs of DFT beams, we form 16 beam pairs as  $(1,3),(2,4),\dots, (15,1),(16,2)$ respectively. 
 The criteria for choosing the auxiliary beam pair is the pair which gives the largest average power out of all the probed beam pairs.
\\
 \figurename{\;\ref{fig:histogram}} shows the convergence behavior of SAGE. By properly initializing SAGE with coarse estimation of spatial frequency $\hat{\mu}_r$, integer delay $\hat{\tau}_i$ assuming $\tau_1 = 0$ and $\hat{\alpha}_r=0$, the maximum number of iterations that SAGE needs to converge is 4 considering $\Gamma = 10^{-3}$. 70\% of all channel realizations, SAGE took 3 iterations to converge to the global optimum.
\begin{figure}[!t]
	\centering
	\includegraphics[width=0.7\linewidth]{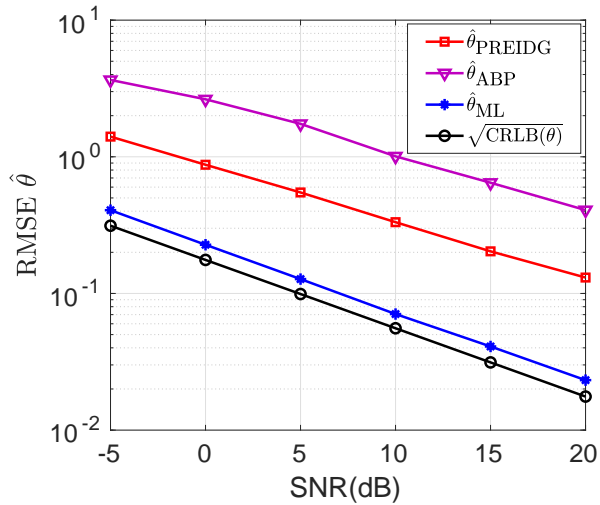}
	\caption{Performance comparison of two-stage and ABP algorithm for LOS AoD only.}
	\label{fig:th}
\end{figure} 
\\
In \figurename{\;\ref{fig:th}}, the two-stage estimation algorithm is evaluated and compared with ABP for LOS AoD estimation, given the aforementioned scenario. Simulation results based on $10$ thousands of channel realizations show that the proposed PREIDG method performs better than ABP. After using coarse estimation based on PREIDG, as an ad-hoc estimation to initialize the SAGE algorithm gives the improved ML performance which nearly satisfies the theoretical bound. 
\begin{figure}[!t]
	\centering
	\includegraphics[width=0.7\linewidth]{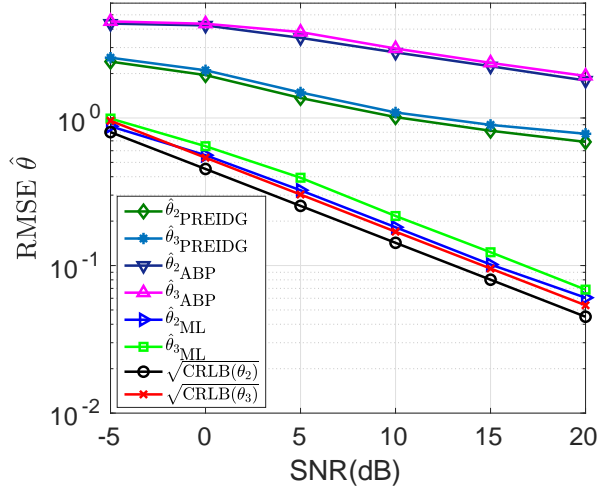}
	\caption{Performance comparison of two-stage and ABP algorithm for NLOS AoDs only.}
	\label{fig:thnlos}
\end{figure} 
\\
The AoDs performance of the two-stage method for NLOS paths are compared with ABP and is assessed with theoretical CRLB in \figurename{\;\ref{fig:thnlos}}.  The performance of PREIDG still performs better than ABP for NLOS paths. The performance of the ML approach performs better than both PREIDG and ABP method. The ML method approaches the CRLB closely.  
\begin{figure}[!t]
	\centering
	\includegraphics[width=0.7\linewidth]{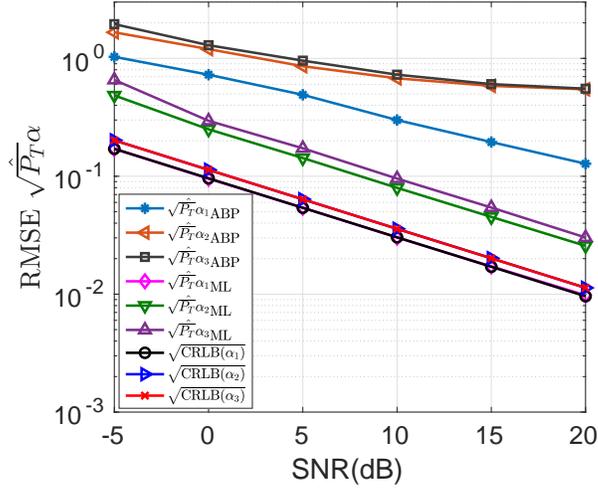}
	\caption{Performance comparison of two-stage and ABP algorithm for $\alpha_r$.}
	\label{fig:alpha_log}
\end{figure} 
\\
\figurename{\;\ref{fig:alpha_log}} shows the performance comparison of complex path gain, $\hat{\sqrt{P_T}\alpha_r}$ via two-stage estimation and ABP method. The performance of the two-stage estimation method for $\hat{\sqrt{P_T}\alpha_r}$ performs better compared to the ABP approach. For the LOS, the two-stage estimation algorithm achieved CRLB because of $\alpha_1 =1$,  while for NLOS paths estimates are not good because of the fact that the model order estimation does not always detect all the NLOS paths. 
\begin{figure}[!t]
	\centering
	\includegraphics[width=0.7\linewidth]{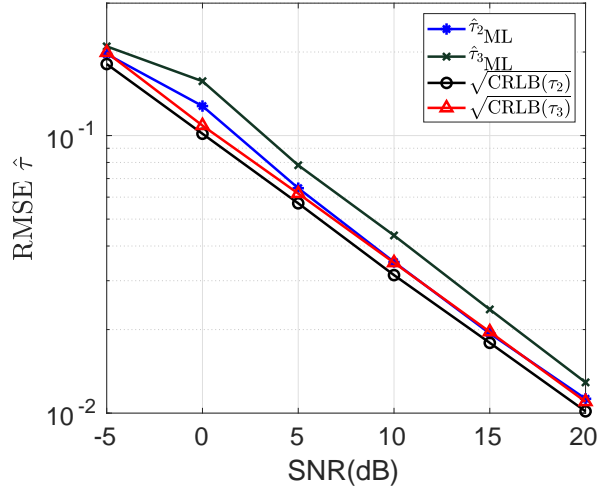}
	\caption{Performance comparison of two-stage algorithm for $\tau_r$.}
	\label{fig:tau_log}
\end{figure} 
\\
Finally in \figurename{\;\ref{fig:tau_log}}, the delay time in fractions of symbol period are estimated for NLOS paths assuming $\tau_1 = 0$ for LOS path. The ABP algorithm is unable to estimate the delay time. The two-stage estimation algorithm performs efficiently and achieves the theoretical bound closely.    \\
Combining both coarse estimation based on PREIDG and using it as an initial guess for SAGE, enhances the estimation accuracy of  $\theta_r$, $\sqrt{P_T}\alpha_r$ and $\tau_r$ drastically, especially in the low SNR regime, which is an important design goal for 5G mmWave systems.

\section{Conclusion}
In this paper, a novel two-stage estimation method is proposed for channel parameter estimation. The coarse estimation is achieved by the novel proposed PREIDG method which is based on interpolation with a fixed LUT. In a second stage, the SAGE algorithm is applied to refine the estimates of the spatial frequency $\hat{\mu}_r$, obtain the complex path gain $\hat{\sqrt{P_T}\alpha}_r$ and, the non-integer delay $\hat{\tau}_r$, for every path. \\
The two-stage estimation method used DFT beams for the estimation of $\mu_r$, $\sqrt{P_T}\alpha_r$ and $\tau_r$ which is efficiently implemented using a Butler matrix in the analog domain and avoid the use of adaptive phase shifters. The proposed two-stage method remarkably reduces the implementation complexity in the analog domain and improves the estimation accuracy and energy efficiency, especially in the low SNR regime which is interesting for 5G mmWave systems. \\
We have derived Cram\'er-Rao lower bound (CRLB) on estimation uncertainty for the spatial frequencies, complex path gains and the delay time between the line of sight and non-line of sight paths.\\
Through simulations, we analyzed and compared the performance of our proposed algorithm with ABP, which proved that the proposed two-stage estimation algorithm has lower implementation complexity and efficient estimation performance. \\
Although the method has been described in detail only for single antenna UE's, it can be extended to multi-antenna UE's, two-dimensional arrays at the BS and usage of orthogonal polarization. 

\appendix
\section{Derivation of ML estimates} \label{Derivation_ML}
The log-likelihood function used in \eqref{costfunction} is
\begin{equation}
\ell(\B{X}_r;\bof{\eta}_r)=\ln \left(\frac{1}{(\pi\beta_r\sigma_n^2)^{ML}}\exp\left(\frac{-\lvert\lvert \B{X}_r - \B{S}_r(\bof{\eta}_r)\rvert\rvert^2_\text{F}}{\beta_r\sigma_n^2}\right)\right).
\end{equation}
By taking the expectation with respect to $\B{X}_r$, we get
\begin{align}
\mathbb{E}_{\B{X}_r}\left[\ell (\B{X}_r;\bof{\eta}_r)|\B{Y};\hat{\bof{\eta}}\right] = -ML \ln \left(\pi \beta_r \sigma_n^2\right) -  \nonumber \\
\frac{1}{\beta_r\sigma_n^2}\left(\R{tr}\left\{\hat{\B{X}}_r^\R{H}\hat{\B{X}}_r
\right\}- \R{tr}\left\{\B{S}_r^\R{H}(\bof{\eta}_r)\hat{\B{X}}_r\right\}- \right. \nonumber \\
\left. \R{tr}\left\{\B{S}_r^\R{T}(\bof{\eta}_r)\hat{\B{X}}_r^\ast\right\} + \| \B{S}_r(\bof{\eta}_r)\|^2_\R{F}	\right).      \label{final_cost}
\end{align}
by simplifying \eqref{final_cost} using the following 
\begin{align}
\| \B{S}_r^\R{T}(\bof{\eta}_r)\|^2_\R{F} = P_T \alpha_r\alpha_r^\ast \R{tr\left\{\B{C}^\R{H}(\tau_r)\B{A}^\R{H}(\mu_r)\B{A}(\mu_r)\B{C}(\tau_r)\right\}}, \label{valueofS}
\end{align} 
by putting \eqref{valueofS} into \eqref{final_cost} can lead us to the following concentrated cost function
\begin{equation}
\hat{\bof{\eta}}_r = \arg\max_{\bof{\eta}_r} \,\B{\Lambda}_r(\bof{\eta}_r),
\end{equation}
where the concentrated $\B{\Lambda}_r(\bof{\eta}_r)$ is represented as 
\begin{align}
\B{\Lambda}_r(\bof{\eta}_r) = \left( \sqrt{P_T}\alpha_r^\ast \R{tr}\left\{\B{C}^\R{H}(\tau_r)\B{A}^\R{H}(\mu_r)\hat{\B{X}}_r\right\} + \right. \nonumber \\
\left. \sqrt{P_T}\alpha\R{tr}\left\{\B{C}^\R{T}(\tau_r)\B{A}^\R{T}(\mu_r)\hat{\B{X}}_r^\ast\right\}- \right. \nonumber \\
\left.  \sqrt{P_T} \alpha_r\alpha_r^\ast \R{tr}\left\{\B{C}^\R{H}(\tau_r)\B{A}^\R{H}(\mu_r)\B{A}(\mu_r)\B{C}(\tau_r)\right\}
\right).
\end{align}
Taking the derivative of $\bof{\Lambda}_r(\bof{\eta}_r)$ with respect to $\sqrt{P_T}\alpha_r^\ast$ and setting $\frac{\partial \bof{\Lambda}_r(\bof{\eta}_r)}{\partial \sqrt{P_T}\alpha_r^\ast}=0$, leads to
\begin{equation}
\hat{\sqrt{P_T}\alpha}_r = \frac{\R{tr}\left\{\B{C}^\R{H}(\hat{\tau}_r)\B{A}^\R{H}(\hat{\mu}_r)\hat{\B{X}}_r\right\}}{ \R{tr}\{\B{C}^\R{H}(\hat{\tau}_r)\B{A}^\R{H}(\hat{\mu}_r)\B{A}(\hat{\mu}_r)\B{C}(\hat{\tau}_r)\}},
\end{equation}
and finally to 
\begin{align}
\left(\hat{\tau}_r,\hat{\mu}_r\right)= \arg \max_{\tau_r, \mu_r} \left\{\frac{ \left|\R{tr}\left\{\B{C}^\R{H}(\tau_r)\B{A}^\R{H}(\mu_r)\B{\hat{X}}_r\right\}\right|^2  }{\beta_r\sigma_n^2 \; \R{tr}\{\B{C}^\R{H}(\tau_r)\B{A}^\R{H}(\mu_r)\B{A}(\mu_r)\B{C}(\tau_r)\}}\right\}.
\end{align}
which is used to iteratively and sequentially solve for $\hat{\tau}_r$ and $\hat{\mu}_r$ in \eqref{tauestimator} and \eqref{muestimator}.
\section{Entries of the FIM $\B{F}(\bof{\eta})$} \label{Derivation_CRLB}
The entries of the block matrices of the FIM \eqref{FIM} are derived as 
\begin{align}
[\B{F}_{\R{Re}\{\bof{\alpha}\}\R{Re}\{\bof{\alpha}\}}]_{i j}&=  \frac{P_T}{\sigma_{n}^2}2\R{Re}\left(\R{tr}\left\{\sum_{r=1}^{R}\frac{\partial\alpha_r^\ast}{\partial\R{Re}\{\alpha_i\}}\B{C}^\R{H}(\tau_r)\B{A}^\R{H}(\mu_r) \right.\right. \nonumber\\
&\left.\left. \sum_{r=1}^{R}\frac{\partial\alpha_r}{\partial\R{Re}\{\alpha_j\}}\B{A}(\mu_r)\B{C}(\tau_r)\right\}\right),
\end{align}
\begin{align}
[\B{F}_{\R{Re}\{\bof{\alpha}\}\R{Im}\{\bof{\alpha}\}}]_{i j}&=  \frac{P_T}{\sigma_{n}^2}2\R{Re}\left(\R{tr}\left\{\sum_{r=1}^{R}\frac{\partial\alpha_r^\ast}{\partial\R{Re}\{\alpha_i\}}\B{C}^\R{H}(\tau_r)\B{A}^\R{H}(\mu_r) \right.\right.\nonumber\\
&\left.\left.\sum_{r=1}^{R}\frac{\partial\alpha_r}{\partial\R{Im}\{\alpha_j\}}\B{A}(\mu_r)\B{C}(\tau_r)\right\}\right),
\end{align}
\begin{align}
[\B{F}_{\R{Re}\{\bof{\alpha}\}\bof{\mu}}]_{i j}&=  \frac{P_T}{\sigma_{n}^2}2\R{Re}\left(\R{tr}\left\{\sum_{r=1}^{R}\frac{\partial\alpha_r^\ast}{\partial\R{Re}\{\alpha_i\}}\B{C}^\R{H}(\tau_r)\B{A}^\R{H}(\mu_r) \right.\right.\nonumber\\
&\left.\left.\sum_{r=1}^{R}\alpha_r\frac{\partial\B{A}(\mu_r)}{\partial\mu_j}\B{C}(\tau_r)\right\}\right),
\end{align}
\begin{align}
[\B{F}_{\R{Re}\{\bof{\alpha}\}\bof{\tau}}]_{i j}&=  \frac{P_T}{\sigma_{n}^2}2\R{Re}\left(\R{tr}\left\{\sum_{r=1}^{R}\frac{\partial\alpha_r^\ast}{\partial\R{Re}\{\alpha_i\}}\B{C}^\R{H}(\tau_r)\B{A}^\R{H}(\mu_r)\right.\right.\nonumber \\
&\left.\left.\sum_{r=1}^{R}\alpha_r\B{A}(\mu_r)\frac{\partial\B{C}(\tau_r)}{\partial\tau_j}\right\}\right), 
\end{align}
\begin{align}
[\B{F}_{\R{Im}\{\bof{\alpha}\}\R{Im}\{\bof{\alpha}\}}]_{i j}&=  \frac{P_T}{\sigma_{n}^2}2\R{Re}\left(\R{tr}\left\{\sum_{r=1}^{R}\frac{\partial\alpha_r^\ast}{\partial\R{Im}\{\alpha_i\}}\B{C}^\R{H}(\tau_r)\B{A}^\R{H}(\mu_r) \right.\right. \nonumber \\
&\left.\left.\sum_{r=1}^{R}\frac{\partial\alpha_r}{\partial\R{Im}\{\alpha_j\}}\B{A}(\mu_r)\B{C}(\tau_r)\right\}\right),
\end{align}
\begin{align}
[\B{F}_{\R{Im}\{\bof{\alpha}\}\bof{\mu}}]_{i j}&=  \frac{P_T}{\sigma_{n}^2}2\R{Re}\left(\R{tr}\left\{\sum_{r=1}^{R}\frac{\partial\alpha_r^\ast}{\partial\R{Im}\{\alpha_i\}}\B{C}^\R{H}(\tau_r)\B{A}^\R{H}(\mu_r)\right.\right. \nonumber \\
&\left.\left.\sum_{r=1}^{R}\alpha_r\frac{\partial\B{A}(\mu_r)}{\partial\mu_j}\B{C}(\tau_r)\right\}\right),
\end{align}
\begin{align}
[\B{F}_{\R{Im}\{\bof{\alpha}\}\bof{\tau}}]_{i j}&=  \frac{P_T}{\sigma_{n}^2}2\R{Re}\left(\R{tr}\left\{\sum_{r=1}^{R}\frac{\partial\alpha_r^\ast}{\partial\R{Im}\{\alpha_i\}}\B{C}^\R{H}(\tau_r)\B{A}^\R{H}(\mu_r)\right.\right. \nonumber \\
&\left.\left.\sum_{r=1}^{R}\alpha_r\B{A}(\mu_r)\frac{\partial\B{C}(\tau_r)}{\partial\tau_j}\right\}\right), 
\end{align}
\begin{align}
[\B{F}_{\bof{\mu\mu}}]_{i j}&=  \frac{P_T}{\sigma_{n}^2}2\R{Re}\left(\R{tr}\left\{\sum_{r=1}^{R}\alpha_r^\ast\B{C}^\R{H}(\tau_r)\frac{\partial\B{A}^\R{H}(\mu_r)}{\partial\mu_i} \right.\right. \nonumber \\
&\left.\left.\sum_{r=1}^{R}\alpha_r\frac{\B{A}(\mu_r)}{\partial\mu_j}\B{C}(\tau_r)\right\}\right),
\end{align}
\begin{align}
[\B{F}_{\bof{\mu\tau}}]_{i j}&=  \frac{P_T}{\sigma_{n}^2}2\R{Re}\left(\R{tr}\left\{\sum_{r=1}^{R}\alpha_r^\ast\B{C}^\R{H}(\tau_r)\frac{\partial\B{A}^\R{H}(\mu_r)}{\partial\mu_i} \right.\right.\nonumber \\
&\left.\left.\sum_{r=1}^{R}\alpha_r\B{A}(\mu_r)\frac{\partial\B{C}(\tau_r)}{\partial\tau_j}\right\}\right),
\end{align}
\begin{align}
[\B{F}_{\bof{\tau\tau}}]_{i j}&=  \frac{P_T}{\sigma_{n}^2}2\R{Re}\left(\R{tr}\left\{\sum_{r=1}^{R}\alpha_r^\ast\frac{\partial\B{C}^\R{H}(\tau_r)}{\partial\tau_i}\B{A}^\R{H}(\mu_r) \right.\right. \nonumber \\
&\left.\left.\sum_{r=1}^{R}\alpha_r\B{A}(\mu_r)\frac{\partial\B{C}(\tau_r)}{\partial\tau_j}\right\}\right).
\end{align}
The partial derivative of $\alpha_r$ and $\alpha_r^\ast$ is calculated as
\begin{align}
\frac{\partial\alpha_r}{\partial\R{Re}\{\alpha_i\}} = \frac{\partial\alpha_r^\ast}{\partial\R{Re}\{\alpha_i\}}=
\begin{cases}
1 \,\, \text{if}\,(r=i) \\
0 \,\, \text{if}\,(r\ne i) ,
\end{cases} \\
\frac{\partial\alpha_r}{\partial\R{Im}\{\alpha_i\}} =- \frac{\partial\alpha_r^\ast}{\partial\R{Im}\{\alpha_i\}}=
\begin{cases}
j \,\, \text{if}\,(r=i) \\
0 \,\, \text{if}\,(r\ne i) .
\end{cases}
\end{align}
The partial derivative of $\B{A}(\mu_r)$ is calculated as
\begin{align}
\frac{\partial \B{A}(\mu_r)}{\partial\mu_r} = \R{diag}\{\B{a}^{\prime\R{H}}(\mu_r)\B{w}(\Phi_k)\}_{k=0}^{M-1},
\end{align}
with
\begin{align}
\B{a}^\prime(\mu_r) = \left[0,\,\,\,\; (-j)\text{e}^{-j\mu_r},\dots,(-j(M-1))\text{e}^{-j(M-1)\mu_r}\right]^\R{T}.
\end{align}
The partial derivative of the sequence $c(t)$ with respect to $\tau_r$ is calculated as
\begin{align}
c(t) = \sum_{n = -\infty}^{+\infty}c(n) \,h(t-nT_s), 
\end{align}
where $h(t)$ is the raised cosine (RC) pulse
\begin{align}
h(t) = \frac{\sin\, (\pi\frac{t}{T_s})}{\pi\frac{t}{T_s}}\frac{\cos\,(\rho\pi\frac{t}{T_s})}{1 - (2\rho\frac{t}{T_s})^2}, 
\end{align}
where $\rho \in [0,1]$, represents the roll-off factor. The delayed sequence by $\tau_r$ is represented as
\begin{equation}
c(t-\tau_r) = \sum_{n=-\infty}^{+\infty} c(n)h(t-nT_s-\tau_r).
\end{equation}
The partial derivative with respect to $\tau_r$ can be written as
\begin{align}
\frac{\partial c(t-\tau_r)}{\partial \tau_r} = \sum_{n=-\infty}^{+\infty}c(n)\frac{\partial h(t-nT_s-\tau_r)}{\partial \tau_r} \nonumber \\
= -\sum_{n=-\infty}^{+\infty} \left(c(n)\,\,\frac{\partial h (\tilde{t})}{\partial \tilde{t}}\,\, \Bigg |_{\tilde{t}=t-nT_s-\tau_r}\right).
\end{align}

\section*{References}
\bibliographystyle{elsarticle-num}
\bibliography{./elsevier_revised}

\end{document}